# The key role of antibonding electron transfer in surface chemisorption and heterogeneous catalysis

Liping Yu, Qimin Yan, and Adrienn Ruzsinszky

Department of Physics, Temple University, Philadelphia, Pennsylvania 19122, USA

## Abstract

The description of the chemical bond between a solid surface and an atom or a molecule is the fundamental basis for understanding surface reactivity and catalysis. Despite considerable research efforts, the physics that rules the strength of such chemical bonds remains elusive, especially on semiconductor surfaces. Widespread understandings are mostly based on the degree of filling of antibonding surface-adsorbate states that weaken the surface adsorption. The unoccupied antibonding surface-adsorbate states are often considered to have no effects on surface bonding. Here, we show that the energy levels of unoccupied antibonding surface-adsorbate states relative to the Fermi-level play a critical role in determining the trends in variations of surface adsorption energies. The electrons that would occupy those high-energy antibonding states are transferred to the Fermi-level, leading to an energy gain that largely controls surface bonding. To illustrate this picture, as a validating case, we study the hydrogen evolution reaction (HER) catalyzed by $MoS_2$ from density functional theory calculations. We find that the majority of antibonding surface-hydrogen states are positioned well above the Fermi-energy. A clear linear relationship between the energy gain from antibonding electron transfer and the adsorption energy is identified for hydrogen binds to either molybdenum or sulfur atoms at different sites. The antibonding-electron transfer energy can thus serve as a primary catalytic activity descriptor. The emerging picture identifies the origin of HER on $MoS_2$, which is related to the empty in-gap states induced by sulfur vacancies or edges. Under this picture, the effects of surface inhomogeneity (e.g., defects, step edges) on surface bonding strength can be understood. This antibonding electron transfer picture also offers a physically different explanation for the well-known *d*-band theory for hydrogen adsorption on transition metal surfaces. The results provide guidelines for understanding and optimizing catalyst performance and designing new solid catalysts.

## Introduction

A key step towards the rational design of new catalysts is to identify active sites and activity descriptors[1-7]. Not all sites at a catalyst surface are catalytically active. The Sabatier principle[8] correlates active sites to the strength of surface-adsorbate interaction, which should be neither too strong nor too weak. Although the interaction strength can be measured by the adsorption energy calculated from first-principles, the identification of the physical and chemical factors that control the surface-adsorbate interaction is challenging. The catalytic surfaces vary with different crystallographic orientations and they are often non-uniform, containing defects, steps and kinks, and corners. These factors as well as different chemical species of surface atoms bonded by the adsorbate often contribute differently to the surface adsorption strength and catalytic activity. Widespread models such as the Newns-Anderson model[9] and the d-band centers[10-13] for transition-metal surfaces, the $e_g$-orbital filling for transition-metal oxide perovskites[14], the lowest unoccupied states for transition metal dichalcogenides[15,16] have been successful in describing the variations of adsorption energy and reactivity on certain well-defined crystallographic planes. However, these models as well as the phenomenological Blyholder model[17] hold only for certain groups of catalysts and are challenged by complex inhomogeneous surface geometries[18,19]. It is therefore desirable to develop a general picture that enables us to understand the microscopic electronic structure origins of surface adsorption and distill useful design principles.

Figure 1 shows the schematic diagram for a surface-adsorbate interaction. When an adsorbate binds to a solid surface, the overlap of their electronic states leads to the formation of bonding and antibonding states. (Note not all surface states actively interact with the adsorbate states and form into bonding and antibonding states. For simplicity, hereafter we call those surface states participating in bonding-antibonding as "active surface states" and those that do not participate in bonding-antibonding as "non-active surface states"). The bonding states are positioned well below the Fermi-energy and are fully occupied[8]. As we shall be showing below, the antibonding states are mostly positioned well above the Fermi-level. The electrons that would be populated into those high-energy antibonding states through orbital interaction are transferred into the Fermi level underneath them[20]. This is the key difference to normal molecular bonds, where no such electron transfer exists.

In this picture, one can write the adsorption energy in the simple form:

$$E_{\text{ad}} = n_\sigma^* \Delta_a - n_\sigma \Delta_b - n_\sigma^{*\prime} \Delta_r \tag{1}$$

where $\Delta_a$ and $\Delta_b$ are, respectively, the antibonding and bonding energy shifts from the midpoint between the active surface-states center ($E_X$) and the adsorbate-states center ($E_A$). $\Delta_r$ is the distance from the antibonding energy center ($E_\sigma^*$) to the Fermi energy ($E_F$). $n_\sigma$, $n_\sigma^*$, and $n_\sigma^{*\prime}$ are the bonding occupancy, the antibonding occupancy before antibonding electron transfer, and the number of antibonding electrons transferred to the Fermi level, respectively. The first two terms in Eq.1 are the energy contributions from antibonding and bonding as in molecules. The third term is the energy gain from antibonding electron transfer.

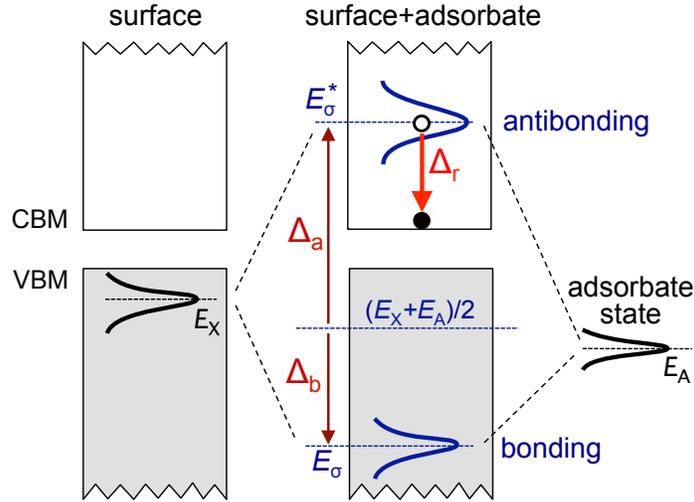

**Figure 1| Schematic diagram for the surface-adsorbate interaction.** The $E_X$ is the center of gravity (in energy) of the surface electronic states that participate in bondin-antibonding with the adsorbate states centered at $E_A$. The $E_\sigma^*$ and $E_\sigma$ are the antibonding and bonding energy centers, respectively, which can be directly calculated from the crystal orbital Hamilton population (COHP) analysis (see Methods section). $\Delta_a$ and $\Delta_b$ are the shifts of the antibonding and bonding energy centers from ½($E_X+E_A$), respectively. The $\Delta_r$ is the distance from the antibonding energy center ($E_\sigma^*$) to the Fermi-energy ($E_F$) of the system after adsorption. The important features are: (i) the majority of antibonding states are often positioned well above the conduction band minimum (CBM), (ii) the number of antibonding electrons transferred to the CBM is often significant and can be as many as the adsorbate electrons, and (iii) the bonding states are fully occupied and the occupation number can be considered the same as that of the surface electrons participating in bonding-antibonding.

Two controlling parameters for bonding on surfaces immediately emerge from this picture. One is active surface-state center $E_X$; the other is the lowest unoccupied state of the surface before adsorption or the Fermi-energy ($E_F$) of the surface after adsorption. For a given type of adsorbate states centered at $E_A$, $E_X$ acts as an input parameter, which determines $E_\sigma^*$ and $E_\sigma$ and hence also $\Delta_a$ and $\Delta_b$ in Eq.1. The $E_F$ and $E_\sigma^*$ define $\Delta_r$. Generally, $E_X$ varies as an adsorbate binds to different surface atoms or the same atom at different surface sites. It implies that the local structure effects on surface bonding manifests themselves in variations of $E_X$.

The main obstacle that impedes progress in revealing trends of chemisorption is how to identify active surface states. Not all surface states are active. In fact, most surface states are non-active, i.e., they do not form bonding-antibonding states with the adsorbate. Active and non-active surface states are mixed together and it is very difficult to disentangle them. Moreover, note that surface band states spread continuously over a wide energy range. The majority of active surface states often fall in one or more narrow sub energy windows without clear boundaries. Such sub energy-window can vary significantly from one surface to another not only in its position (in energy) but also in its size. Even on the same crystallographic surface, the energy windows of active surface states can also differ significantly at different adsorption sites such as defects and edges. In this sense, $E_X$ is not well defined and it is very challenging to quantify it. So neither are $\Delta_a$ and $\Delta_b$ as shown in Fig.1.

Here we circumvent this obstacle to study trends of chemisorption bonds by the antibonding electron transfer energy $\Delta_r$, i.e., $E_\sigma^* - E_F$. Thanks to the crystal orbital Hamilton population (COHP) analysis[21,22], $E_\sigma^*$ and $E_\sigma$ are well defined and can be computed from first principles (see Methods section). Unlike $E_X$, $\Delta_r$ is well defined and calculable. On the other hand, since $E_\sigma^*$ is an output resulting from the interaction between active surface states ($E_X$) and adsorbate states ($E_A$), $E_\sigma^*$ varies with $E_X$. It means that $\Delta_r$ can also capture the local structure effects on chemisorption as well as $E_X$ does. We shall be demonstrating below that $\Delta_r$ can serve as primary descriptor for trends of bonding on surfaces. The proposed method is general and can be applied to any chemical systems with various structural features (e.g., defect, impurities, and edges) that are accessible to the DFT modeling.

## Results

**Role of the antibonding electron transfer**

The role of antibonding electron transfer in revealing trends of chemisorption can be seen from Eq. 1. When antibonding states are as fully occupied as bonding states (i.e. $n_\sigma^* = n_\sigma$), the net bonding-antibonding contribution to the adsorption energy is positive since $\Delta_a > \Delta_b$ according to molecular orbital theory[8]. It means that the surface-adsorbate interaction would always be repulsive if there was no counteracting contribution from antibonding electron transfer. The antibonding electron transfer hence is the only source that can render the adsorbate-surface interaction attractive. This is often the case when a closed-shell adsorbate (e.g., $H_2$ and He) binds to a semiconductor surface. When the antibonding surface-adsorbate states are partially filled (i.e., $n_\sigma^* < n_\sigma$), the bonding-energy contribution is partially compensated by the antibonding-energy contribution. The antibonding electron transfer counteracts this compensation and thus often largely determines the trends in variations of adsorption energies at different sites.

To illustrate this quantitatively, we study the binding strength of hydrogen adsorbed on molybdenum disulfide ($MoS_2$) from density functional theory calculations. Details of calculations are given in the Methods section and the supercell structure models are given in the Supplementary Figure 1. $MoS_2$ is a promising nonprecious HER catalyst[23]. Several methods have been proposed to improve HER, including creating sulfur-vacancies[24], producing edges sites[3,25], making metallic 1T phase[26,27] and molecular structures[28], and applying strains[29]. Yet, the microscopic activity origin and the underlying design principles for controlling the strength of hydrogen adsorption are far from being well understood.

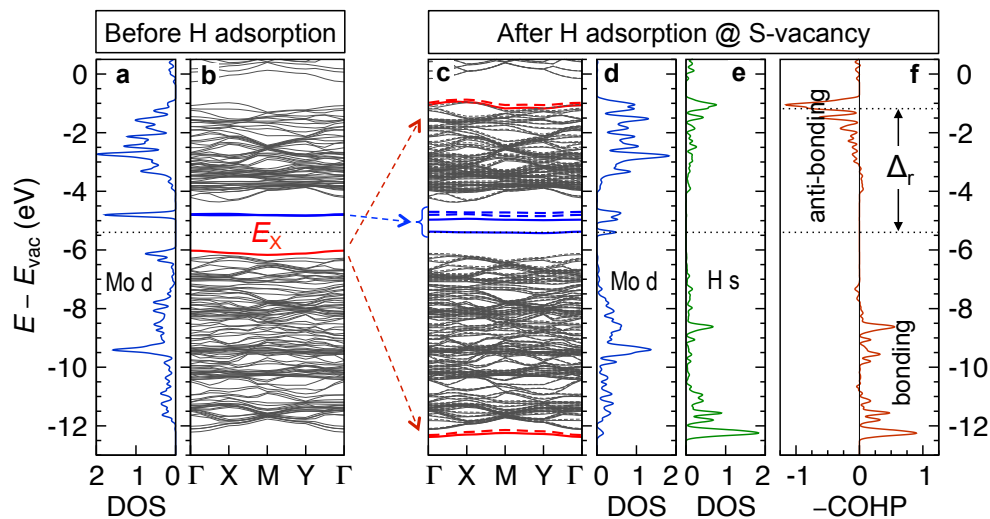

**Figure 2 | Band structure, projected density of states (DOS), and COHP bonding analysis.**
**a**, the DOS projected to the Mo d orbital to which hydrogen will bind later. **b,** the band structures of $MoS_2$ with 3.125% sulfur-vacancies. The in-gap states induced by the S-vacancy are highlighted in colors red (occupied) and blue (unoccupied). **c**, Same as **b** but with hydrogen adsorbed at one S-vacancy. Solid (dashed) lines are for spin-up (spin-down). **d-e**, the DOS projected to Mo d and H 1s, which are bonded to each other. **f**, the COHP for the Mo-H bond. Bonding interactions to the right and antibonding interactions to the left. The dotted lines in **a-f** are the Fermi-energy (-5.4eV) after hydrogen adsorption. The prominent features are: (i) the $E_X$ state in **b** disappears in **c**, splitting into bonding and antibonding states, (ii) the electron that would occupy the Mo-H antibonding state is transferred to the lowest unoccupied in-gap states, and (iii) $\Delta_r$ is large and no Mo-H antibonding states appear near and below the Fermi-level.

Figure 2 illustrates the antibonding electron transfer process in $MoS_2$ with hydrogen atom adsorbed on a sulfur vacancy site. The chemical bonding-antibonding information is analyzed through the COHP projected from plan-wave basis sets[21,22]. The results resemble the simple picture of Fig. 1 very well. Removing one S from MoS$_2$ breaks three bonds and leaves two electrons at the vacancy site. These three dangling bonds form three in-gap states that are mainly composed of Mo d orbitals (Fig.2a). The lowest one is doubly occupied and lies near the valence band maximum (VBM). The other two are unoccupied and positioned about 1 eV below the conduction band minimum (CBM) (Fig.2b). When hydrogen atom is placed at the center of the vacancy, the hydrogen 1s orbital strongly interacts with the doubly occupied S-vacancy states. This interaction forms a bonding state located about 6 eV below the VBM and an antibonding state positioned about 3 eV above the CBM (Fig.2cdef). Two of the three electrons (one from hydrogen and two from the occupied in-gap state) fill the bonding state; the third electron, which would occupy the antibonding-state lying high in the conduction bands, is transferred to the lowest unoccupied in-gap states that are also induced by the S-vacancy (Fig.2b). Clearly seen is that almost all antibonding hydrogen-suface states are positioned well above the Fermi-level (Fig.2f). The large $\Delta_r$ (3.64 eV) originates from the low-energy empty in-gap states induced by the S-vacancy. The energy gain from this antibonding electron transfer is thus significant.

**Linear dependence of adsorption energy on antibonding electron transfer**

Figure 3a summarizes our calculated adsorption energies ($\Delta E_H$) for hydrogen adsorption on various MoS$_2$ surfaces and edges, and the corresponding antibonding electron transfer energy $\Delta_r$ obtained from the COHP analysis. Detailed electronic band structures, density of states, and COHP analysis are shown in Supplementary Figures 2-4. A clear linear relationship is found between $\Delta E_H$ and $\Delta_r$. The slope of this linear relationship (solid line) is −1.18, which is close to −1, meaning that $\Delta E_H$ variation is mostly (but not completely) reflected in the $\Delta_r$ variation. The larger the $\Delta_r$, the smaller the adsorption energy. The linear relationship for H-Mo bindings (cases E-G) and H-S bindings (cases A-D as shown in Fig.3a) has a slope of -2.24 and -1.12, respectively. It suggests that the bonding and antibonding energy (first two terms in Eq.1) also contribute more to the $\Delta E_H$ variation in the latter (i.e., the cases of Mo-S bindings).

The linear relationship between $\Delta E_H$ and $\Delta_r$ is understandable. For H-S bindings, Fig. 3b shows that the bonding and antibonding energy centers $E_\sigma$ and $E_\sigma^*$ increase as one goes from case A to case D, whereas their differences, $E_\sigma^* - E_\sigma$, are almost invariant. The former is expected since the active surface state center $E_X$ increases as both $E_\sigma$ and $E_\sigma^*$ increase from case A to case D: the metallic 1T-phase (case B) has a Fermi-energy higher than the VBM of 1H-phae (case A), and the armchair and zigzag edges (cases C and D) have in-gap edge states (due to dangling bonds) higher in energy than the Fermi-energy of case B. The invariance in $E_\sigma^* - E_\sigma$ can be attributed to the fact that the hydrogen atom is coordinated to a single sulfur atom with an almost same H-S bond length in all cases A-D. Given that $n_\sigma^{*\prime}$, $n_\sigma^*$, and $n_\sigma$ are generally much the same under different H-S bindings, the first two terms in Eq. 1 are also largely same. Therefore, the $\Delta E_H$ depends almost exclusively on $\Delta_r$ in a linear decreasing way. For H-Mo bindings, $E_\sigma$ and $E_\sigma^*$ vary differently. So do the $\Delta_a$, $\Delta_b$ and $E_\sigma^* - E_\sigma$. In cases E, F, and G, hydrogen atom binds to one, two, and three Mo atoms, respectively. The more Mo atoms bonded to hydrogen, the more degree of their orbital overlap. It is thus expected that $E_\sigma^* - E_\sigma$ follows the order: E (12.65 eV) < F (13.52 eV) < G (13.85 eV), along which the hydrogen's coordination number increases. Note that the $\Delta_r$ follows in a different order: E (3.40 eV) < G (3.64 eV) < F (4.29 eV), due to different Fermi levels associated with different in-gap states. Since $\Delta_r$ varies more significantly than $E_\sigma^* - E_\sigma$ and the bonding-antibonding contributions to $\Delta E_H$ compensate each other, the $\Delta E_H$ follows the trend of $\Delta_r$ for H-Mo bindings.

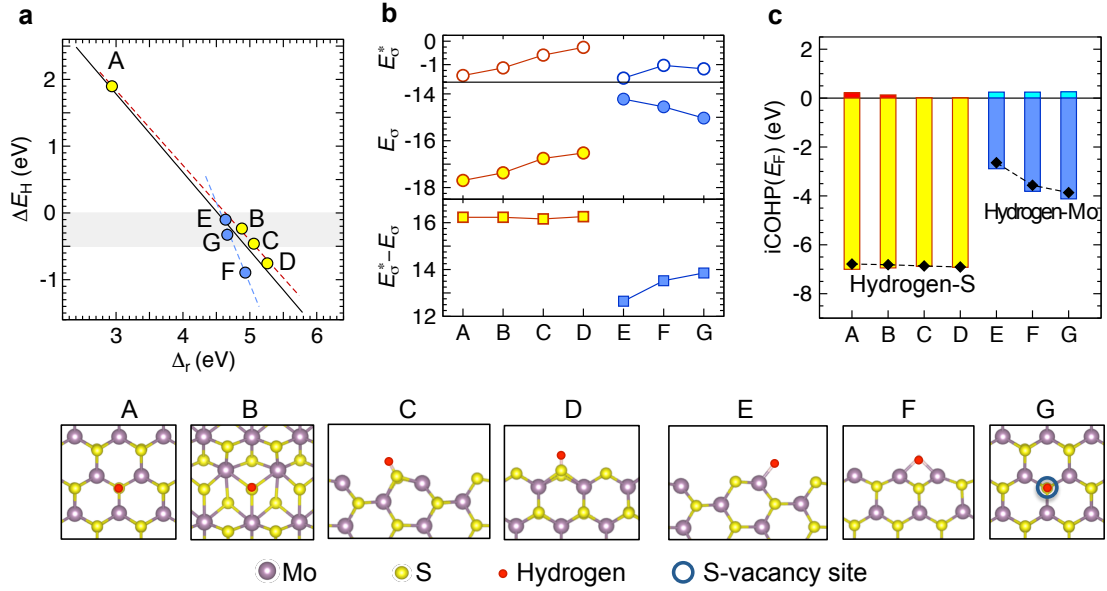

**Figure 3 | Role of antibonding electron transfer on hydrogen adsorption on MoS$_2$ surfaces and edges.** **a,** Calculated adsorption energy ($\Delta E_H$) for hydrogen atom in seven different configurations as shown in panels A-G. The black solid line is the fitting to all data points. The red and blue dashed lines are the fitting to points A-D and points E-G, respectively. **b,** Calculated antibonding state center ($E_\sigma^*$), bonding-state center ($E_\sigma$), and their differences. The energies are in eV and relative to the vacuum level, which is set to zero in the plots. **c,** the integrated COHP (iCOHP) up to the Fermi-level. The yellow and blue bars (below zero) are the bonding contributions; the red and cyan bars (above zero) are the antibonding contributions. The black diamond symbols are the total bonding-antibonding contributions. Panels A-D are the configurations with hydrogen binds to sulfur atoms. Panels E-G are the configurations with hydrogen binds to Mo atoms. Except that B is for MoS$_2$ in the 1T phase, all others are for MoS$_2$ in the 1H phase. The grey shaded area in **b** marks the Gibbs free energy of hydrogen adsorption ranging from about -0.25 eV to 0.25 eV, where the sites can be catalytically active.

**Occupancy of bonding-antibonding states**

The above implies that considering only the degree of filling of antibonding and bonding states is insufficient for describing the trends of surface adsorption energies[29,30]. The contributions from those occupied bonding and antibonding states can be estimated from the energy integral of the corresponding COHP up to the Fermi-energy[22]. For H-S bindings, Fig.3c shows that the bonding contributions (yellow bars) are almost constant among all four cases considered. Their antibonding energy contributions (red bars) are fractional (<4%) compared to the bonding-energy contributions. Neither of them alone nor their total (black symbols) can explain the $\Delta E_H$ variation shown in Fig.3b. For H-Mo bindings, Fig.3c shows that the antibonding contributions (cyan bars) are also fractional, less than 10% of their corresponding bonding contributions (blue bars). Both the bonding and antibonding contributions alone and their total follow the order E<F<G, inconsistent with the $\Delta E_H$ variation order (E<G<F).

**Coordination number of the adsorbate**

Particularly worthy of mention here is the role of the coordination number of the adsorbate. (Note it is not the coordination number of the surface atoms bonded to the adsorbate.[7,19]) Fig.3c shows the net energy contribution from occupied antibonding-bonding states increases with the hydrogen's coordination number. The Fermi-energy and its associated $\Delta_r$ do not necessarily follow this trend (Fig.3a). For the same coordination number, such net antibonding-bonding energy contribution is almost unaffected by whether the surface atoms bonded by the adsorbate have dangling bonds, whether they reside in different crystal structures, or whether they are located at the surface planes with different crystallographic orientations. However, it does not mean that dangling bonds, surface plane orientations, and crystal structures have no influence on surface binding. In fact, they still affect surface binding through $\Delta_r$, because they often result in different empty in-gap states or CBM that yield different $\Delta_r$ (Figs. 2-3 and Supplementary Figs. 2-4). Therefore, the coordination number of the adsorbate is an important factor for controlling the surface bonding strength.

**The antibonding-electron transfer model versus the *d*-band model**

Here we compare the antibonding electron transfer picture with the well-known *d*-band model[10,11]. The *d*-band model is remarkably useful in understanding bond formation and trends in reactivity among the transition metals. It approximately describes the surface-adsorbate interaction by the *d*-band center of a transition metal surface: the lower the *d* states are in energy relative to the Fermi-level, the more the antibonding states are filled and the weaker the binding.

Table 1 summarizes our calculated $\Delta E_H$, *d*-band center $\varepsilon_d$, and $\Delta_r$ for hydrogen adsorbed on the (111) surfaces of metals Ni, Cu, and Ag. As expected from the *d*-band model, $\Delta E_H$ is found to increase as $\varepsilon_d$ becomes more negative (i.e., deeper from the $E_F$) from Ni to Cu to Ag, consistent with previous calculations[31]. Within the antibonding-electron transfer picture, Table 1 and supplementary Fig. 5 indicates that a more negative $\varepsilon_d$ corresponds to a smaller $\Delta_r$, which leads to a larger $\Delta E_H$ and a weaker binding. Hence, the antibonding electron transfer model is consistent with the *d*-band model in describing the trends of chemisorption on the transition metal surfaces.

**Table 1. Calculated hydrogen adsorption properties on the (111) surfaces of metals Ni, Cu, and Ag.** The lower the d-band center, the smaller the antibonding electron transfer energy.

|  | Adsorption energy $\Delta E_H$ (eV) | *d*-band center $\varepsilon_d$ (eV) | Antibonding electron transfer energy $\Delta_r$ (eV) |
|---|---|---|---|
| Ni (111) | −0.61 | −1.64 | 2.58 |
| Cu(111) | −0.31 | −3.00 | 1.78 |
| Ag(111) | 0.16 | −4.63 | 1.67 |

However, the antibonding electron transfer model is physically different from the *d*-band model. The former is based on the energy levels of *unoccupied* anitbonding surface-adsorbate states, whereas the latter is rationalized from the *occupancy* of antibonding states. In the former, the $\Delta_r$ results from the surface states ($E_X$) that actually participate in bonding-antibonding with adsorbate states. In the latter, the $\varepsilon_d$ refers to the center of *d*-states that not all of them participate in bonding-antibonding with the adsorbate states. In fact, usually only partial *d*-states (e.g., $d_{xz}$ and $d_{yz}$) within a certain energy range participate in bonding-antibonding. In addition, $\Delta_r$ contains the contribution from the interaction between the transition metal *s* states and the adsorbate states, which is ignored in the *d*-band model. Therefore, the variation in $E_X$ (and also adsorption energy) is expected to manifest itself better in the $\Delta_r$ variation than that in the $\varepsilon_d$ variation.

The antibonding-electron transfer model is more general than the *d*-band model. The d-band model holds mainly for transition metals or transition metal compounds with the adsorbate directly bonded to the transition metal atom(s). It cannot explain the chemisorption trend revealed here for MoS$_2$ systems, because (i) H does not always bind to the Mo atom(s), and (ii) the occupancy of those antibonding surface-H states is too low to affect the strength of hydrogen binding.

## Discussion

The emerging picture of antibonding-electron transfer identifies not only the active sites for HER on MoS$_2$ but also their microscopic origin. Consistent with experiments, we find that edges[3,25,32] and S-vacancies[24,29] in 1H-MoS$_2$ and the basal plane of 1T-MoS$_2$ [26,27] are catalytically active, whereas the basal plane of 1H-MoS$_2$ is catalytically inert. The HER at those active sites originates from the emergence of low-energy empty states that lead to larger antibonding electron transfer energy $\Delta_r$. The occupancy of antibonding states turns out too small to determine the strength of hydrogen binding on MoS$_2$. The clear linear relationship demonstrated between the surface adsorption energy and the $\Delta_r$ indicates that $\Delta_r$ is a good descriptor for trends of surface bonding and reaction activity. This $\Delta_r$ descriptor is well defined and calculable from COHP analysis and DFT. It can capture inhomogeneous local surface structure effects through the antibonding energy center $E_\sigma^*$ (indicative of $E_X$) and the CBM.

The above results provide guidelines on how to optimize the catalytic performance of a given catalyst. Usually, surface catalysis is not optimized, because the surface-adsorbate binding is either too strong or too weak. Here, the results indicate that the binding strength strongly depends on $\Delta_r$. Experimentally, $\Delta_r$ can be tuned by changing the CBM through strain engineering[29,33,34], defect engineering[4,5,35], nanostructure engineering[5], substrate engineering[36], alloying[12], etc. Specifically, if an adsorbate does not bind or binds too weakly to a surface, one can strengthen the binding by applying stains that lower the CBM, making nanostructures with empty in-gap edge states, or creating acceptor-like defects with empty in-gap states, or using metal substrates with a Fermi-energy lower than the CBM of the catalyst. If the surface-adsorbate binding is too strong, it can be weakened by the strains that increase the CBM, or the metal substrates with a Fermi-energy higher than the CBM, etc. Note we presume that the aforementioned methods induce less significant change in $E_\sigma^*$ than that in the CBM for a given type of catalysts, which might often be the case as suggested from Fig.3. Nevertheless, one can use the approach proposed here to study the effects of strains, defects, and substrates on $\Delta_r$ from first principles, providing guidance on how to apply corresponding experimental methods to improve the catalytic performance of a given catalyst.

The physical picture presented here can also be applied to explore trends of adsorption on chemically different surfaces. The concepts emerging from this picture are not restricted to a specific system. In principle, they are applicable for any semiconductor or metallic surface interacting with any atom or molecule. For perfect crystalline surfaces from same chemical group in the same crystal structure, the chemical trends of adsorption on $\Delta_r$ might be more easily seen since fewer parameters are involved. For example, for transition metal dichalcogenides (MX$_2$) in the 1T or 1H structure, a positive but very scattered linear relationship was found recently between the free energy of hydrogen adsorption ($\Delta G_H$) and the CBM[16]. Within the picture presented here, this linearity is expected since a lower CBM often leads to a larger $\Delta_r$ and a smaller $\Delta G_H$ for a fixed $E_\sigma^*$. The scattering feature may be largely attributed to variations of $E_\sigma^*$. In addition, the CBM of MX$_2$ mostly follows the trend that it increases as X goes from S to Te, or M goes to the right (bottom) in the transition-metal rows (columns) of the periodic table[15,37-39]. Based on the physical picture presented here, $\Delta G_H$ would also follow this chemical trend if the role of $E_\sigma^*$ is relatively less significant. This trend thus suggests that groups 4-5 MX$_2$ (M=Ti, Zr, Hf; V, Nb, Ta) are most likely HER active on their basal planes as opposed to MoS$_2$, consistent with recent experimental finding[15,40].

## Methods

All calculations were performed using density functional theory (DFT) and the plane-wave projector augmented-wave (PAW) method [41,42] as implemented in the VASP code[43]. The SCAN (strongly constrained and appropriately normed) meta-generalized gradient approximation[44] was used. SCAN predicts geometries and energies of diversely bonded molecules and materials (including covalent, metallic, ionic, hydrogen, and van der Waals (vdW) bonds) with remarkable accuracy[45,46]. It also captures the intermediate-range vdW interaction about right, which is missing in PBE-GGA and in most hybrid GGAs. (The intermediate range is roughly the distance between nearest-neighbor atoms at equilibrium.) For example, SCAN provides a significant improvement over PBE-GGA in computing the absolute and relative stability of main group compounds, approaching the chemical accuracy of 0.04 eV/atom in formation enthalpy and providing reliable structure selection as an indicator of accurate relative stability.[46] For layered materials (including MoS$_2$ considered here), the SCAN with a long range vdW correction[47] yields excellent interlayer binding energies and spacings, as wells as intra-layer lattice constants.

A kinetic energy cutoff of 400 eV and 500 eV was used for the plane-wave expansion, respectively, for MoS$_2$ systems and metal systems (Ni, Cu, and Ag). All atoms were fully relaxed until their atomic forces were less than 0.02 eVÅ$^{-1}$. The electronic calculations for the system with hydrogen adsorption were carried out using spin-polarized approaches. The dipole correction[48] was applied in order to remove the artificial dipole interaction caused by using the slab supercell method for surface calculations. During structural relaxations, a k-point mesh of 2×2×1 and 6×6×1 was used, respectively, for MoS$_2$ systems and metal systems.

The hydrogen adsorption energy on various MoS$_2$ surfaces, edges, and S-vacancy defects (Fig.3a and Table 1) was calculated from $\Delta E_\mathrm{H} = E_{\mathrm{MoS}_2+\mathrm{H}} - E_{\mathrm{MoS}_2} - 1/2 E_{\mathrm{H}_2}$, where $E_{\mathrm{MoS}_2+\mathrm{H}}$ and $E_{\mathrm{MoS}_2}$ are the DFT total energy of the fully relaxed MoS$_2$ supercell structure with and without H adsorption, respectively. $E_{\mathrm{H}_2}$ is the DFT total energy of hydrogen molecule. The supercell models for the basal planes of MoS$_2$ in both 1T and 1H phases contain 16 Mo atoms and 32 S atoms. The hydrogen adsorption at the edge site was modeled in a supercell containing 30 Mo atoms and 60 atoms. For face-centered cubic metals (Ni, Cu, and Ag), a four-atomic-layer (111) slab structure was constructed with 16 metal atoms in each layer (See Supplementary Fig.5a). The H was adsorbed on the three-fold hollow site of metal surfaces. In all supercells, a 15 Å vacuum space was inserted in the c-direction to create surfaces for studying adsorption properties. The detailed information of these DFT-relaxed structures is shown in Supplementary Fig. 1 and Supplementary Data 1.

The chemical bonding was investigated using the COHP analysis as projected from DFT plan-wave basis sets[21,22]. The COHP is a partitioning of the band-structure energy in terms of orbital-pair contributions. The interaction between two orbitals, $\phi_u$ and $\phi_v$, centered at neighboring atoms, is described by their Hamiltonian matrix element $H_{uv} = \langle \phi_u | H | \phi_v \rangle$. The COHP is defined as the multiplication of this matrix element with the corresponding densities-of-states matrix. It serves as a quantitative measure of bonding strength because the product either lowers (bonding) or raises (antibonding) the DFT band-structure energy. Here, we define the centers in energy of bonding and antibonding adsorbate-surface states relative to the vacuum level, respectively, as $E_\sigma = \int_{-\infty}^{\infty} \mathrm{COHP}_-(E) E dE / \int_{-\infty}^{\infty} \mathrm{COHP}_-(E) dE$, and $E_\sigma^* = \int_{-\infty}^{\infty} \mathrm{COHP}_+(E) E dE / \int_{-\infty}^{\infty} \mathrm{COHP}_+(E) dE$. The $\mathrm{COHP}_+(E)$ and $\mathrm{COHP}_-(E)$ are the positive COHP due to antibonding and the negative COHP due to bonding, respectively. The COHP(E) are all aligned to the vacuum level and so are the calculated $E_\sigma^*$ and $E_\sigma$ shown in Fig.3b. The integrated COHP shown in Fig.3c is given as $\mathrm{iCOHP}(E_\mathrm{F}) = \int_{-\infty}^{E_\mathrm{F}} \mathrm{COHP}(E) dE$. The antibonding electron transfer energy ($\Delta_\mathrm{r}$) (Fig.3a and Table 1) is defined as $\Delta_\mathrm{r} = \int_{E_\mathrm{F}}^{\infty} \mathrm{COHP}_-(E) dE - E_\mathrm{F}$, where $E_\mathrm{F}$ is the Fermi-level (i.e., the VBM) of the system after hydrogen adsorption.

## Acknowledgements

We thank John P. Perdew for valuable scientific discussions and comments on the manuscript. This research was supported as part of the Center for the Computational Design of Functional Layered Materials (CCDM), an Energy Frontier Research Center funded by the U.S. Department of Energy (DOE), Office of Science, Basic Energy Sciences (BES), under Award #DE-SC0012575. This research used resources of the National Energy Research Scientific Computing Center, a DOE Office of Science User Facility supported by the Office of Science of the U.S. Department of Energy under Contract No. DE-AC02-05CH11231. This research was also supported in part by the National Science Foundation through major research instrumentation grant number CNS-09-58854.

## Author contributions

L.Y. designed the project, developed the model, performed the calculations, and wrote the manuscript. Q.Y. and A.R. provided support and contributed to analyzing the results and writing the manuscript.

## Competing interests

The authors declare no competing financial interests.

## Corresponding author

Correspondence to Liping Yu (E-mail: yuliping@gmail.com ) or Qimin Yan (E-mail: qiminyan@temple.edu ).

**Data availability**

The authors declare that the data supporting the findings of this study are available within the paper and its Supplementary figure and data files.

# Supplementary Information

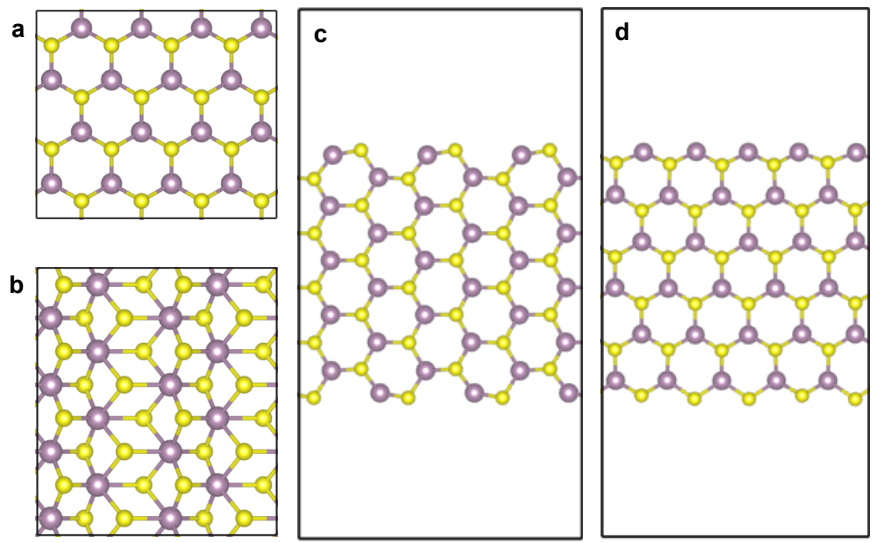

**Supplementary Figure 1 | Supercell structures adopted for calculating hydrogen adsorption energies on MoS$_2$.** **a**,**b**, in-plane crystal structures of MoS$_2$ in 1H and 1T phase, respectively. Hydrogen atom is adsorbed atop the sulfur atom in the basal plane. **c,d,** MoS$_2$ nanoribbons with armchair-shaped and zigzag-shaped edges, respectively. Hydrogen binds to either Mo or S atom at each edge. Sulfur and Mo atoms are shown in color yellow and grey respectively.

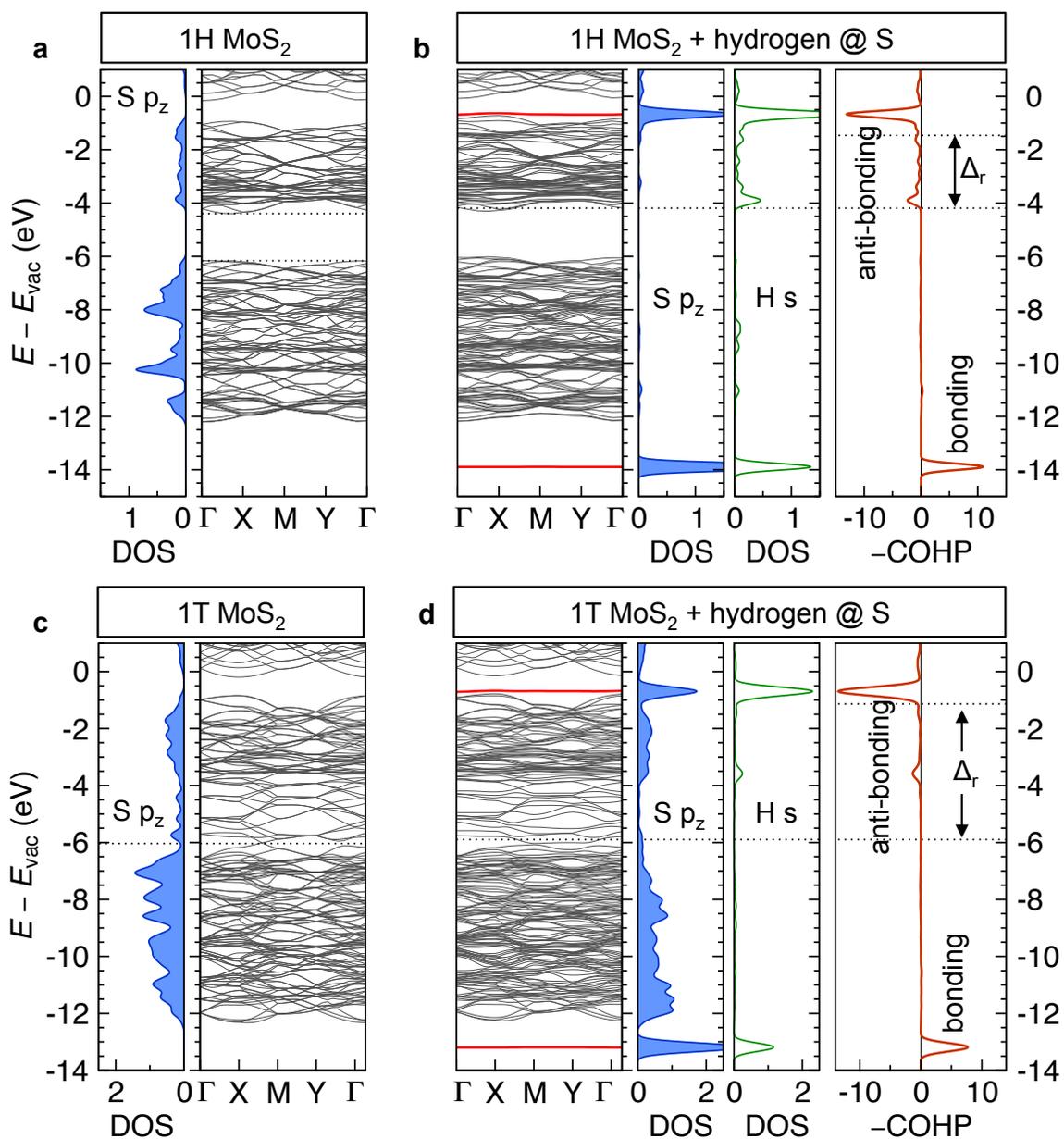

**Supplementary Figure 2 | Band structure, projected density of states, and COHP analysis. a,b,** hydrogen adsorption on the basal plane of $MoS_2$ in 1H structure. **c,d,** hydrogen adsorption on the basal plane of $MoS_2$ in 1T structure. Dotted horizontal lines mark the VBM and CBM in **a,** and the Fermi-level in the band structure and DOS plots of **b-d**. The prominent bonding and antibonding bands are highlighted in color red in the band structures shown in **b** and **d**. The antibonding states below the Fermi-level are ignorable.

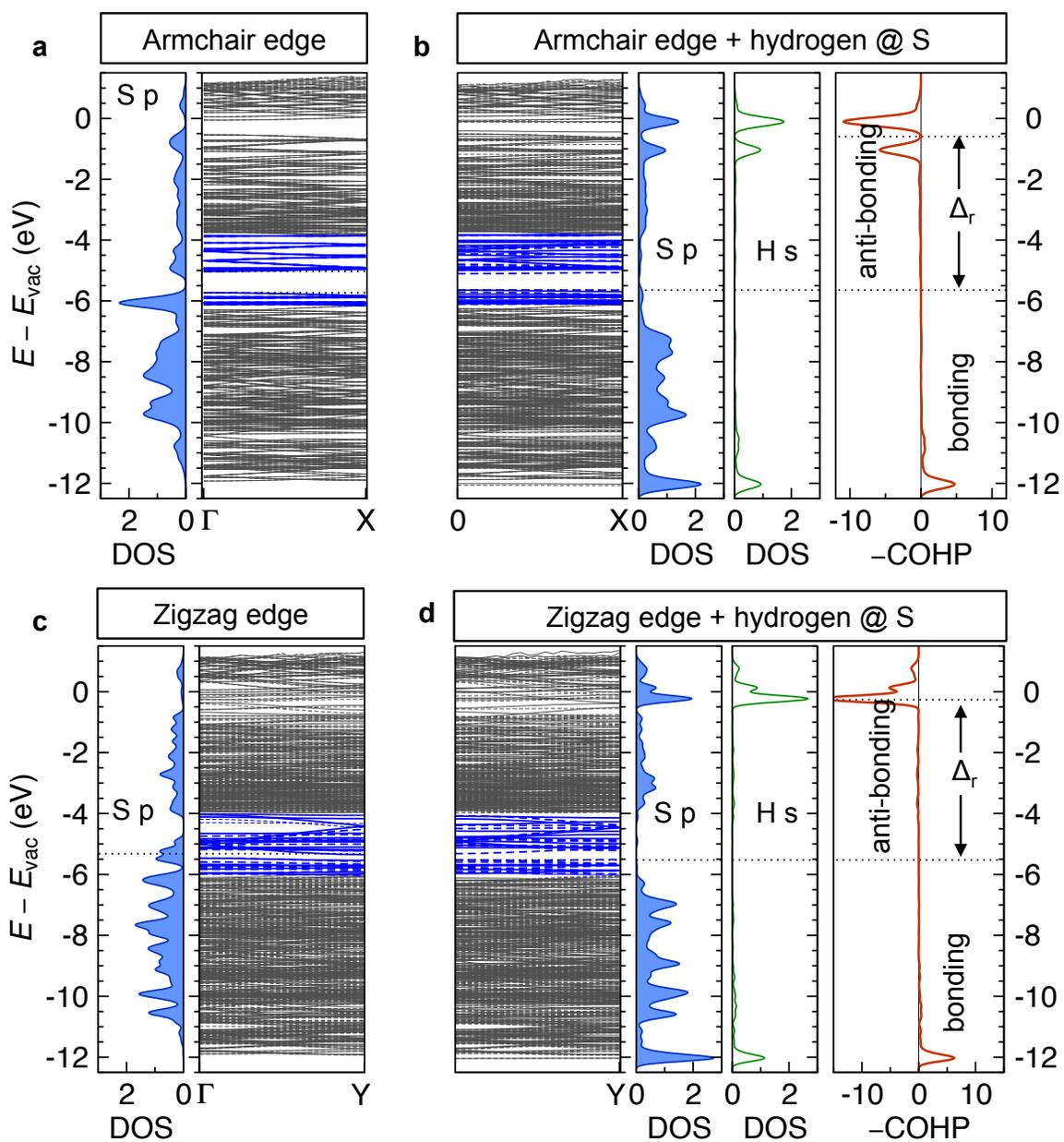

**Supplementary Figure 3 | Band structure, projected density of states, and COHP analysis. a,b,** hydrogen adsorbed on the sulfur atom of one armchair-shaped edge. **c,d,** hydrogen adsorbed on the sulfur atom of one zigzag-shaped edge. The edge-induced in-gap-states bands are highlighted in color blue in their band structures. Dotted horizontal lines mark the Fermi-level. The antibonding states are negligible below the Fermi-level.

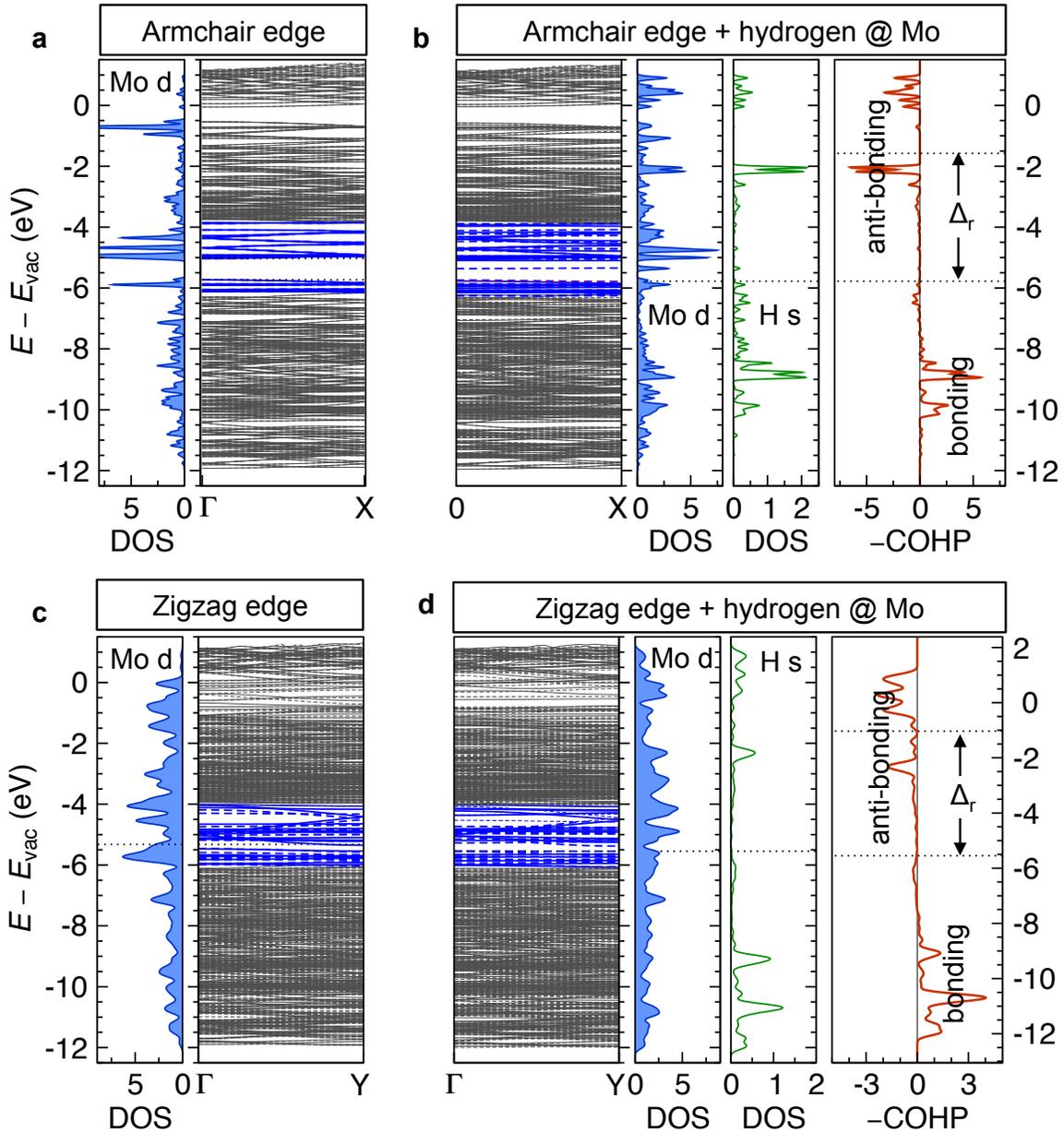

**Supplementary Figure 4 | Band structure, projected density of states, and COHP analysis. a,b,** hydrogen adsorbed on the Mo atom of one armchair-shaped edge. **c,d,** hydrogen adsorbed on the Mo atom of one zigzag-shaped edge. The edge-induced in-gap-states bands are highlighted in color blue in their band structures. Dotted horizontal lines mark the Fermi-level. Fractional antibonding states are found below the Fermi-level.

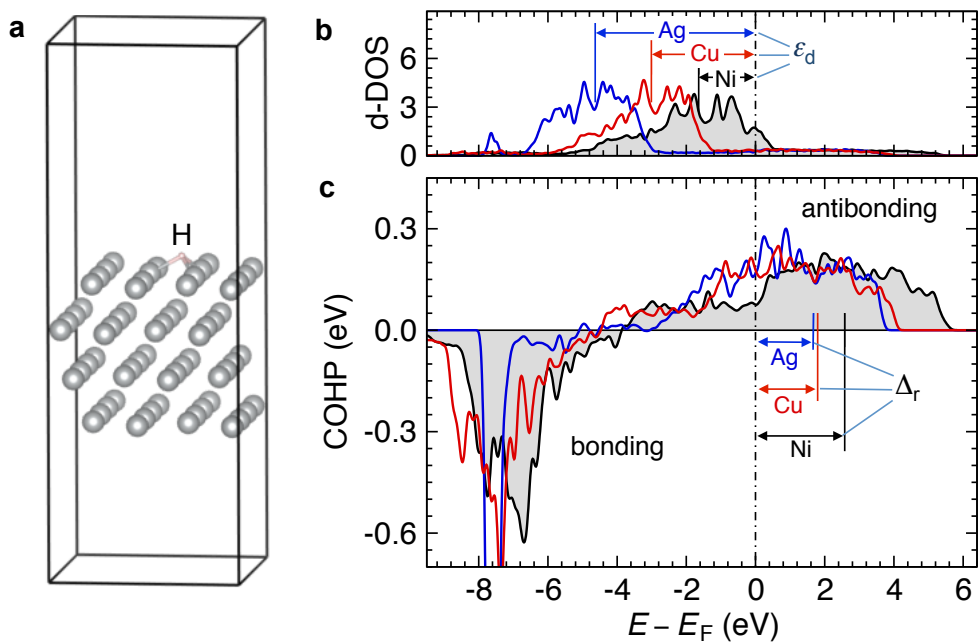

**Supplementary Figure 5 | Projected density of states and COHP analysis for hydrogen adsorption on metals Ni, Cu, and Ag. a**, the slab supercell structure adopted in the calculations. **b**, the density of states projected to the d states of the metal atom that is directly bonded by the hydrogen atom. **c**, the corresponding COHP analysis. The vertical bars in **b** and **c** mark the positions of the d states and the unoccupied antibonding COHP, respectively.